\documentclass[12pt]{iopart}

\newcommand\newblock{\hskip .11em\@plus.33em\@minus.07em}
\usepackage[square,sort,comma,numbers]{natbib}
\usepackage{graphicx,color} 
\usepackage{subcaption}
\usepackage{url}
\usepackage{graphicx,color} 
\usepackage{hyphenat} 
\usepackage{hyperref}
\usepackage{float}
\usepackage{tabularx}
\usepackage{multirow}
\usepackage{booktabs}
\usepackage{siunitx}
\usepackage{lineno}
\usepackage{amsfonts}
\usepackage{bm}
\usepackage{tikz}
\usetikzlibrary{shapes.geometric, arrows}
\usepackage{pgfplots,tikz-3dplot}
\usetikzlibrary{calc}
\usepackage{tikz-3dplot}
\usepackage{soul,xcolor} 
\usepackage{graphicx,color} 
\setstcolor{red}

\newcommand{\orcid}[1]{\href{https://orcid.org/#1}{\includegraphics[scale=0.15]{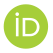}}}

\begin{document}

\title[The impact of local noise recorded at the ET candidate sites]{The impact of local noise recorded at the ET candidate sites on the signal to noise ratio of CBC gravitational wave signals for the ET triangle configuration}

\author{Matteo Di Giovanni$^{1,2,3}$\footnote{matteo.digiovanni@sns.it} \orcid{0000-0003-4049-8336}, Davide Rozza$^{4,5}$ \orcid{0000-0002-7378-6353}, Rosario De Rosa$^{6,7}$ \orcid{0000-0002-4004-947X}, Enrico Calloni$^{6,7}$ \orcid{0000-0003-4819-3297}, Domenico D'Urso$^{8,9,10}$ \orcid{0000-0002-8215-4542}, Luca Naticchioni$^{11}$ \orcid{0000-0003-2918-0730}, Annalisa Allocca$^{6,7}$ \orcid{0000-0002-5288-1351}, Giovanni Luca Cardello$^{8,9,10}$ \orcid{0000-0003-0521-7949}, Alessandro Cardini$^{9}$ \orcid{0000-0002-6649-0298}, Andrea Contu$^{9}$ \orcid{0000-0002-3545-2969}, Giovanni Diaferia$^{12}$ \orcid{0000-0001-9663-0477},Luciano Errico$^{6,7}$ \orcid{0000-0003-2112-0653},Carlo Giunchi$^{10}$ \orcid{0000-0002-0174-324X},Jan Harms$^{13,14}$ \orcid{0000-0002-7332-9806},Irene Molinari$^{12}$ \orcid{0000-0002-8314-1444}, Marco Olivieri$^{12}$ \orcid{0000-0002-7333-8809},Piero Rapagnani$^{3,11}$ \orcid{0000-0002-1865-6126},Valeria Sipala$^{8,9}$ \orcid{0000-0003-3639-8359}Lucia Trozzo$^{7}$ \orcid{0000-0002-8803-6715}, and Fulvio Ricci$^{3,11}$ \orcid{0000-0001-5475-4447}}

\address{$^{1}$ Scuola Normale Superiore, I-56126 Pisa, Italy}
\address{$^{2}$ INFN, Sezione di Pisa, I-56126 Pisa, Italy}
\address{$^{3}$ La Sapienza Università di Roma, I-00185 Roma, Italy}
\address{$^{4}$ INFN, sezione di Milano Bicocca, I-20126, Milano, Italy}
\address{$^{5}$ Università di Milano Bicocca, I-20126, Milano, Italy}
\address{$^{6}$ Universitá Federico II Napoli, I-80126 Napoli, Italy}
\address{$^{7}$ INFN, sezione di Napoli, I-80126 Napoli, Italy}
\address{$^{8}$ Università degli Studi di Sassari, I-07100, Sassari, Italy}
\address{$^{9}$ INFN, sezione di Cagliari, I-09042, Monserrato (Cagliari), Italy}
\address{$^{10}$ INGV, Sezione di Pisa, I-56125, Pisa, Italy}
\address{$^{11}$ INFN, Sezione di Roma, I-00185 Roma, Italy}
\address{$^{12}$ INGV, Sezione di Bologna, I-40127, Bologna, Italy}
\address{$^{13}$ Gran Sasso Science Institute, I-67100, L'Aquila, Italy}
\address{$^{14}$ INFN, Laboratori Nazionali del Gran Sasso, I-67100, Assergi (L'Aquila), Italy}

\vspace{10pt}
\begin{indented}
\item[]\date{\today}
\end{indented}

\begin{abstract}
We present an evaluation of how site dependent noise can affect the signal to noise ratio (SNR) of compact binary coalescence (CBC) signals in the future 3$^{\rm{rd}}$ generation gravitational wave (GW) detector Einstein Telescope (ET). The design of ET is currently pushing the scientific community to study its scientific potential with respect to known, and possibly unexpected, GW signals using its design sensitivity curves. However, local ambient noise may have an impact on the ET sensitivity at low frequency and therefore affect the SNR of CBC signals at low frequency. Therefore, we study the impact of ambient noise on the ET sensitivity curve at the two sites candidate to host ET - Sardinia, in Italy, and the Euregio Meuse-Rhine (EMR) at the Netherlands-Belgium border - and infer the impact on the ET sensitivity curve and how the SNR of CBC signals at low frequencies is affected. We find that Sardinia shows results which are on par, if not  better, than the design case. On the other hand, ambient noise for the current EMR sensitivity curve in Terziet causes a higher degradation of the SNR 
performances. 

\begin{description}
\item[Keywords: ]
Einstein Telescope, Gravitational waves, Newtonian noise, compact binary coalescence
\end{description}
\end{abstract}
\section{Introduction}
\label{sec:Introduction}
With the successful exploitation \citep{gwtc1, gwtc2, gwtc3} of current 2$^{\rm{nd}}$ generation gravitational wave (GW) detectors, namely Advanced Virgo (AdVirgo), located in Italy \citep{aVirgo}, Advanced LIGO Hanford and Livingston (aLIGO), located in the United States \citep{aLIGO} and KAGRA in Japan \citep{kagra}, the scientific community started to thoroughly investigate \cite{Maggiore_2020, coba} the prospects of future 3$^{\rm{rd}}$ generation GW detectors that are expected to start observations in the mid 2030s, i.e., Cosmic Explorer (CE) \cite{CE1, CE2, CE3} and the Einstein Telescope (ET) \cite{et, ET2010, ET2011, ET2020}, with the latter being the subject of this work. ET was first proposed in 2010 and the foreseen improvements with respect to current-generation detectors include the extension of the observation bandwidth from the current limit of about 20\,Hz to 2\,Hz and an improvement of the sensitivity up to a factor 8 across the band covered by current detectors \citep{ET2020}. 

For what concerns the configuration of ET, there are two proposals under consideration. The most recent one is that of a detector network composed by two widely separated L-shaped detectors with 15 km long arms\cite{coba, Iacovelli_2024}. On the other hand, the original project foresees three pairs of nested interferometers arranged in an equilateral triangle (Figure \ref{xylophone}) with the sides 10\,km long \cite{et, ET2010, ET2011, ET2020}. For each interferometer pair, one detector is optimized for low frequencies (2\,Hz $< f <$ 40\,Hz) and the other for high frequencies ($f>$ 40\,Hz). Since recently, moving the lower limit to 3 Hz is being considered as well \citep{ET2020}, but, to be on the conservative side, we keep using the 2 Hz lower limit for this work. In both the 2L and the triangle configurations, ET will be hosted underground, at a currently planned depth between 200\,m and 300\,m to reduce seismic motion at the input of the suspension system of the mirrors and to reduce the impact of atmospheric disturbances \citep{hutt} and Newtonian noise (NN). In this work, since it poses the most challenges from the point of view of environmental noise mitigation, we focus on the ET triangular configuration. 

Generally speaking, the extension of the bandwidth to 2\,Hz and the sharp increase in sensitivity will significantly improve the rate of detected events giving the possibility to issue early warnings for the coalescence of compact objects (CBC) several minutes, if not hours depending on the source, before the merger \cite{Branchesi_2016,Maggiore_2020,Nitz_2021,coba, Hu_2023, refId0}. In fact, with respect to current detectors, CBC signals will spend more time in the ET accessible bandwidth, therefore enabling early detection and sky localization. As it was demonstrated by GW170817 \cite{multim, Branchesi2018, radice, nature, refId0}, this is extremely important in the case of binary neutron star (BNS) mergers. As a matter of fact, joint multimessenger observations with electromagnetic observatories are paramount to exploit the scientific potential of BNS events and shed new light on the internal structure of neutron stars \cite{Maggiore_2020}. In particular, depending on the distance of the source and requiring a signal-to-noise ratio (SNR) $\geq 12$ and a sky localization smaller than $100 \rm\, deg^2$, ET is expected to be able to send early warnings between 1 and 20 hours before the BNS merger \cite{Maggiore_2020, refId0}. The release of an early warning increases the chances of detection for the electromagnetic counterparts. Astronomers would be able to point the telescope in the region of the signal to start the monitoring to obtain pre-merger images, ensuring the coverage of the merger and post merger phases as well. This enables to detect the early electromagnetic emission, which is fundamental to understand the physics of the emission mechanism and the merger remnant \cite{Maggiore_2020, refId0}. Moreover, this class of signals will spend most of their time in the ET bandwidth in the $[2,10]\rm\,Hz$ range, thus making the low frequency sensitivity of ET extremely important for early detection.

The significant increase in sensitivity below $100 \rm\, Hz$ will also allow for more detailed observations about the merger of intermediate mass black holes (IMBH) \cite{Koliopanos:2018sW, mezcua, Maggiore_2020, greene_2020}, i.e. black holes (BH) with masses in the $[10^2, 10^5]\rm\,M_{\bigodot}$ range, of which there was no clear evidence about their existence. This until the third observing run (O3) of the LIGO-Virgo detectors, which observed an event consistent with two BH merging to form a $\simeq 140 \rm M_{\bigodot}$ remnant, providing the first direct evidence of IMBH formation \cite{GW190521}. During the same observing run, a few more marginal candidates for IMBH formation were found, but none sufficiently significant to indicate detection of further IMBH mergers \cite{imbho3,lvkimbh}. The typical merger frequency of this class of astrophysical objects is $<100 \rm\, Hz$ and the signals are expected to spend a maximum of $\mathcal{O}(100\rm\, s)$ in the ET bandwidth with most cycles of their waveform contained in the $[2,10]\rm\,Hz$ band. Currently, GW detectors are limited to BH masses $\mathcal{O}(100\rm\, M_{\bigodot})$ and IMBH merger signals are typically only detectable in very short time intervals of a few milliseconds. On the other hand, ET will open the possibility of detecting these IMBH, studying their formation channels and the possibility that they are the seeds of the supermassive BHs in the center of galaxies \cite{Maggiore_2020, greene_2020}. Furthermore, a precise reconstruction of waveforms for compact objects during inspiral and merger will allow accurate tests of General Relativity. 

Therefore, any degradation with respect to design in the low-frequency sensitivity of ET may significantly hinder the capability of early detections and multimessenger observations for BNS mergers and may reduce the quality of the observations for IMBH \cite{Nitz_2021, coba}. As a consequence, since seismic disturbances, of both natural \cite{higgins1950, ward1966wind, cessaro1994, withers, acernese2004properties, coward2005characterizing, virgo2006, burtin2008spectral, anthony18,   smith, dybing, o3noise, anthony} and anthropogenic \cite{acernese2004properties, virgo2006, saccorotti, piccinini, poli, o3noise} origin, are the main source of noise limiting the detector sensitivity at low frequency and can affect GW data in many ways \citep{virgo2004, virgo2006, virgo2011, cvse, koleyThesis, Fiori2020, o3noise, accadia2010noise, MaEA2016, saulson, beccaria1998relevance, harms}, seismic characterization studies at the candidate sites to host ET are paramount. The goal is to guarantee a suitable environment for this future detector that makes the reaching of its design sensitivity possible \citep{amann} through appropriate design of noise suppression systems. 

\begin{figure}
     \centering
     \includegraphics[width=\linewidth]{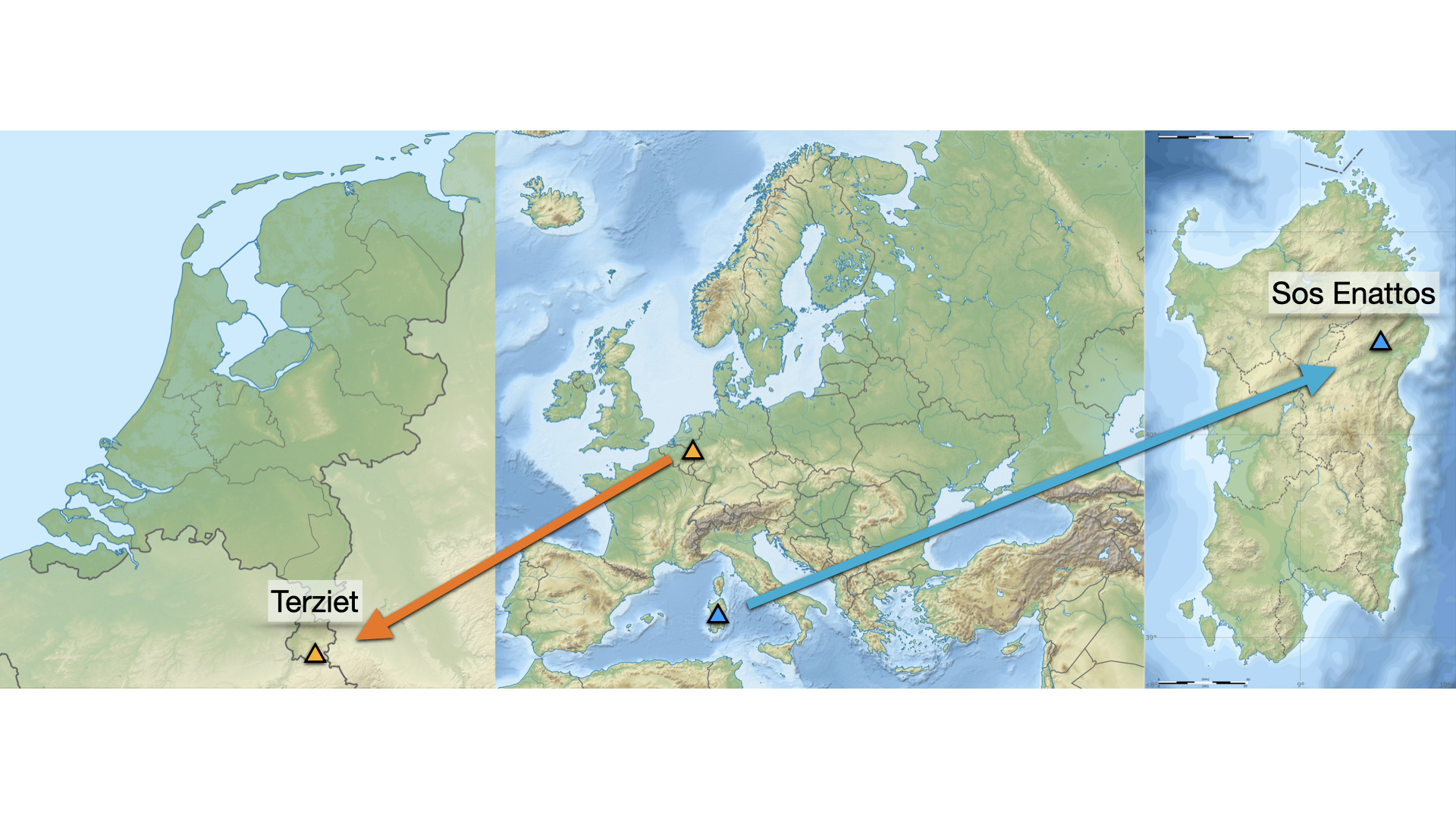}
     \caption{\textit{Map of Europe showing the locations of the candidate sites considered in this work. The village of Terziet in the Netherlands is usually taken as a reference for the EMR candidate site. The maps are taken from \cite{sardiniamap, netherlandsmap, europemap} and modified according to the creative commons license 3.0. The orientation of the triangle icons is not representative of the actual proposed orientations for the ET detector for each site.}}
     \label{fig:map}
 \end{figure}

As of 2024, there are two sites which are officially candidate to host the ET triangle (Figure \ref{fig:map}): the  Euregio Meuse-Rhine (EMR) \cite{ET2011, ET2020}, between Belgium and the Netherlands, and the area surrounding the Sos Enattos former mine in Sardinia (Italy) \cite{ET2011, ET2020, naticchionietal2014, naticchionietal2020, digiovannietal2021, digiovanni2023}.
\begin{center}
    \begin{figure}[t]
\begin{tikzpicture}[scale=1]
    \draw [yshift=5] [line width=4] [blue] (0,0) -- (2.1,3.464) -- (4,0);
    \draw [yshift=0] [xshift=-1.5] [line width=4pt] [orange] (0.15,0) -- (4,0) -- (2.2,3.264);
    \draw [yshift=-4][xshift=-0.5] [line width=4pt] [green] (4,0) -- (0,0) -- (2.15,3.564);
    \coordinate[label={[green]left:ET1}]  (ET1) at (-0.2,0);
    \coordinate[label={[orange]right:ET2}] (ET2) at (4.1,0);
    \coordinate[label={[blue]above:ET3}] (ET3) at (2.1,3.764);
    \coordinate[label=below:{$L=10\rm\, km$}](c) at (2.05,-0.5);
  \end{tikzpicture}
\caption{\textit{Scheme of ET for the triangular configuration. Each detector ET1,2,3 is composed of an interferometer optimized for low-frequency detection and another optimized for high-frequency detection for a total of six.}}
\label{xylophone}
    \end{figure}
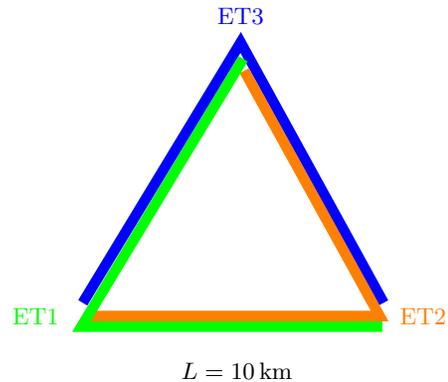
\end{center}
Since 2014, the Sos Enattos \cite{naticchionietal2014, naticchionietal2020, digiovannietal2021, alloccaetal2021, digiovanni2023, saccorotti23} and EMR \cite{koley19, Bader_2022, koley2022surface,vanBeveren_2023} sites have been the subject of characterization studies
aimed at assessing their suitability to host ET. However, except Ref. \cite{alloccaetal2021} and the recent work published by Ref. \cite{janssens2024}, no works are assessing the impact of site dependent noise on the detectability of GW sources. Therefore, the scope of this paper is to provide a first estimation of how site dependent low-frequency noise may affect the SNR of CBC signals after modifying the ET sensitivity curve according to the actual seismic noise measured at each of the aforementioned ET candidate sites. We think that this approach may provide some relevant information for both the site selection process and the design of noise suppression systems for ET. In particular, extending the methodology already shown in Refs.\cite{alloccaetal2021} and \cite{badaracco2019}, we investigate the case of IMBH and BNS focusing on the $[2,10]\rm\, Hz$ range, where ambient noise is more prominent \cite{digiovannietal2021, digiovanni2023, alloccaetal2021} and may affect early warnings. Beyond 10 Hz, noise contributions from the detector hardware dominate the noise budget and, therefore, are out of the scope of our work. As a consequence of the focus in the low frequency range, our goal is not to claim whether a merger will be detected or not (merger frequencies are way beyond the considered frequency range), but only if it will be possible to observe the inspiral preceding the merger for multimessenger purposes. On the other hand, the details of how parameter estimation and other cosmological implications would be affected are beyond the goal of this work and will not be discussed here, but may represent one of the future developments of this study. 

The paper is organized in the following way: in Section \ref{sec:Method} we present the method; Section \ref{sec:Data} talks through the data used for this study; the results are discussed in Sections \ref{sec:curves} and \ref{sec:snr}. Finally, the conclusions are debated in Sections \ref{sec:disc} and \ref{sec:conc}.

\section{Method}
\label{sec:Method}

To estimate the impact of site-dependent noise over the SNR of CBC events, we first calculate how site noise changes the ET design sensitivity curve. For this study we use the design curve already applied for Ref. \cite{coba}. Depending on the frequency band, different contributions affect the ET noise budget curve like seismic, Newtonian, thermal, and quantum noise. Among these noise sources, the dominant contribution in the frequency region between 2 and 10 Hz is the NN. Beyond 10 Hz, other contributions, like quantum and thermal noise from the detector hardware, dominate the noise budget \cite{beker12, beker2015, harms, harms2022}. NN is caused by gravity fluctuations from mass distributions that change around the detector with time. These fluctuations are generated by dynamic variations of atmospheric and ground densities in the vicinity of the test masses of the detectors. Ground density variations are caused by ground motion, i.e., seismic waves, which are the only contribution to NN considered in this work.
Atmospheric NN is not taken into account due to the current lack of dedicated environmental sensors installed at the ET candidate sites.
 
We estimate the NN contribution of the ET sensitivity between 2 Hz and 10 Hz as \cite{harms2022}
\begin{equation}\label{eq:NN}
    \tilde{h}_{\rm{NN}}(f) = \frac{4\pi}{3}G\rho_0\frac{2\sqrt{2}}{L}\frac{1}{(2\pi f)^2}\tilde{x}(f),
\end{equation}
where the frequency ($f$) dependence of the NN budget ($\tilde{h}_{\rm{NN}}$) is related to the gravitational constant ($G$), the density of the ground medium ($\rho_0$), the size of the arm-length of each interferometer ($L$) and the amplitude spectral density (ASD) of the seismic displacement ($\tilde{x}$). Equation \ref{eq:NN} can be derived from 
\begin{equation}\label{eq:deltaacc}
    \delta\vec{a}(\omega) = \frac{4\pi}{3}G\rho_0\left(2\vec{\xi_P}(\omega)-\vec{\xi_S}(\omega)\right),
\end{equation}
which is Equation 5 of \cite{alloccaetal2021} and Equation 6 of \cite{harms2022}, where $\vec\xi_{P,S}$ is the seismic displacement vector for the compressional (P) and shear (S) waves, respectively, and $\omega$ is the angular frequency. Equation \ref{eq:NN} holds for a spherical cavern in an infinite homogeneous medium as it is clearly stated after Equation 6 of Ref.\cite{harms2022}. An even more detailed derivation can be found in Section 3.3 of Ref. \cite{harms}.
In general, equation \ref{eq:NN} is evaluated assuming:
\begin{itemize}
    \item underground seismic spectra representative of body waves only;
    \item one third of the body waves contribution is from compressional waves;
    \item surface waves are negligible at depths of a few hundred meters;
    \item uncorrelated NN on the interferometer test masses.
\end{itemize}
In addtion to that, Equation \ref{eq:NN} does not include any suppression factor from NN mitigation systems.

The ASD of seismic noise ($\tilde{x}$ of Equation \ref{eq:NN}) is evaluated calculating single fast Fourier tranforms (FFT) of the horizontal components of seismic noise, which, contrary to the vertical component, are dominated by body waves waves \cite{IMBERGER2013333, Kulhanek2011}, on $120\rm\,s$ windows. The length of the time window is chosen according to the typical maximum duration of an IMBH signal (Figure \ref{fig:time_spent}) and on the time segment duration proposed for CBC multiband analysis for BNS \cite{Hu_2023}. Each FFT segment is then used to evaluate the probabilistic power spectral density (PPSD) of seismic noise  over 2 years of data from which we infer the percentiles (10th, 50th, and 90th) of the spectral distributions. Under these assumptions, the ET noise budget that we use takes into account the NN evaluated from the seismic noise of each candidate site for the 10th, 50th, and 90th percentile of seismic spectra.
Since a simple NN projection alone does not show the effect of non-stationarity of NN (from now on defined glitchness) we also use a more effective indicator, called the Noise to Target Ratio (NTR) and derived from Ref. \cite{alloccaetal2021} as:
\begin{equation}\label{eq:NTR}
    \rm{NTR} = \sqrt{\frac{1}{\Delta f}\int_{f_1}^{f_2}{df\frac{S_{n}(f)}{S_{n,\rm{ET}}(f)}}},
\end{equation}
where the $S_{n,\rm{ET}}(f)$ is the PSD of the ET design sensitivity, whereas $S_{n}(f)$ is the PSD of the modified ET sensitivity, where the NN is substituted with the contribution of the local noise over the chosen time window and $\Delta f=f_2-f_1$ is the selected bandwidth. This way, the NTR is evaluated time by time and normalised for the comparison of the two candidate sites.

After inferring the modified ET sensitivity curves, we gather the events of interest for our study from astrophysical catalogs and we use PyCBC \cite{pycbc} to generate and inject signals in simulated ET noise. Waveforms are generated using the IMRPhenomD \cite{imrphenomd, imrphenomd2} GW approximant for IMBH and IMRPhenomPv2\_NRTidalv2 \cite{imrphenomp} for BNS, both available in the LIGO-Virgo-KAGRA Algorithm Library (LAL) \cite{lalsuite}. To take into account the antenna pattern of the detector, the ET triangle is simulated considering three properly oriented L-shaped co-located detectors, with an arm length of $10 \rm\, km$ and an arm angle of $60^{\circ}$. The variation of the antenna pattern with time is also taken into account. Since the actual orientation of the ET triangle has not been decided yet, we choose an arbitrary positioning of the detector. For the Sardinia candidate site, ET is assumed to be located at the Sos Enattos mine at $[40.44^{\circ} \rm\,N, 9.44^{\circ} \rm\,E]$, whereas for the EMR site, we have chosen the village of Terziet at $[50.73^{\circ} \rm\,N, 5.91^{\circ} \rm\,E]$, a proposed location for at least one of the ET vertexes, as a reference. For each detector, noise is obtained by generating a noise time series according to the given sensitivity curve and the same sources are taken into account. 
Waveforms are injected into the noise of each detector composing ET (labeled ET1, ET2 and ET3 respectively).

First, we infer the distribution of the SNR of the signals injected in simulated noise obtained using the ET design sensitivity. Signal SNR is calculated as
\begin{equation}
    SNR= \sqrt{MSNR_{\rm{ET1}}^2+MSNR_{\rm{ET2}}^2+MSNR_{\rm{ET3}}^2},
\end{equation}
where $\rm\,MSNR_{\rm{ET1,2,3}}$ are the matched filter SNR calculated in each detector. Then, using the same events, we repeat the procedure for each curve inferred from Equation \ref{eq:NN} using the 10th, 50th and 90th percentile seismic spectra. The new SNR distributions are then compared against the design case used as a benchmark. The distributions of the SNR losses are obtained after calculating, for each event, the ratio between the new SNR and the benchmark SNR. 

To calculate the MSNR, we assume the signal to exhibit the two polarization states\cite{Allen}
\begin{equation}\label{eq:plus}
    h_{+}(t)=-A(t) \frac{1+\cos^2(\iota)}{2}\cos[2\phi_c+2\phi(t)],
\end{equation}
\begin{equation}\label{eq:cross}
    h_{\times}(t)=-A(t) \cos^2(\iota)\sin[2\phi_c+2\phi(t)],
\end{equation}
where A(t) is the amplitude, $\iota$ the inclination of the source, $\phi_c$ the phase of the signal at the chirp and $\phi(t)$ the phase of the signal. Therefore, MSNR is calculated as \cite{Usman_2016}:
\begin{equation}\label{eq:snr1}
    MSNR^2(t) = \frac{(s|h_{cos})^2+(s|h_{sin})^2}{(h_{cos}|h_{cos})}
\end{equation}
with
\begin{equation}\label{eq:snr2}
    (s|h)(t)=4 \mathrm{Re}\int_{f_{\rm{min}}}^{f_{\rm{max}}}\frac{\tilde s(f)\tilde h^*(f)}{S_n(f)}e^{2\pi i ft}df
\end{equation}
where $\tilde s(f)$ is the Fourier Transform (FT) of the data, $\tilde h^*(f)$ is the FT of the template and $h_{cos}$/$h_{sin}$ are the two orthogonal phase components of the template shown in Equations \ref{eq:plus} and \ref{eq:cross} \cite{Allen, Usman_2016}. In our case, $f_{\rm{min}} = 2\rm\, Hz$ and $f_{\rm{max}} = 10\rm\, Hz$ because, as mentioned above, we consider ambient noise only which modifies the ET sensitivity curve not beyond the $[2,10]\rm\, Hz$ range. Moreover, taking into account Equations \ref{eq:snr1} and \ref{eq:snr2} we note that MSNR out of this bandwidth gives the same contribution to the total MSNR in all cases and would not bring relevant information. Therefore we restrict the calculation between 2 Hz and 10 Hz only. We also point out that we assume perfect knowledge of the source, i.e., matched filter is calculated using the same waveform template injected in the data. 
\section{Data}\label{sec:Data}
\subsection{Seismic}\label{subsec:seismic}
Seismic data from the candidate sites come from a set of instruments installed to assess the quality of the sites and for site characterization studies. In particular, data from the Sardinia site come from two seismometers installed in boreholes at $\sim -264$ m and $\sim -252$ m at the two proposed vertex locations for the ET triangle known as P2 and P3. The seismometers are Trillium 120 SPH2 coupled with a Nanometrics Centaur CTR4-6S, 6-ch 24-bit data logger. For what concerns the EMR site, the only available data are from a Streckeisen STS-5A borehole seismometer in Terziet, referred to as TERZ, at $\sim -250$ m. For this reason, we use TERZ as a reference for the EMR candidate site. In both cases, the time period used for the analysis covers two years, from December 21$^{\textrm{st}}$, 2021 to December 20$^{\textrm{th}}$, 2023. This period of 730 days has a duty cycle $>90\%$ at both sites. 
Data from P3 shows a lower noise level with respect to P2 in all the channels by an average factor of around 20\% between 1 and 10 Hz. For this reason, we will compare P2 and TERZ. 
 \begin{figure}
     \centering
     \includegraphics[width=0.5\linewidth]{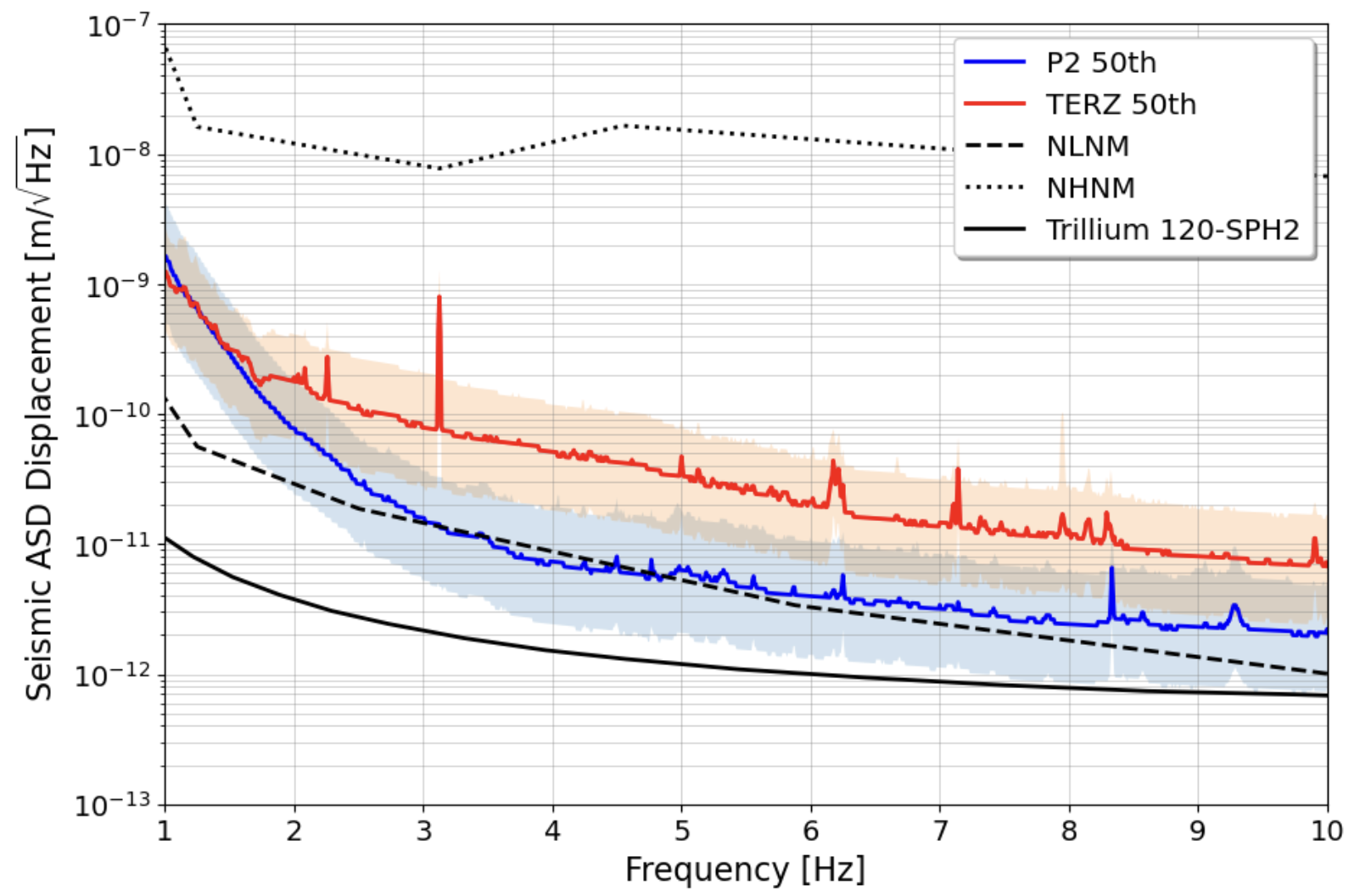}
     \caption{\textit{Seismic ASD displacement for the two candidate sites, TERZ in red and P2 in blue, are reported as a percentile (50th in solid line while 10th and 90th the lower and upper limit of the colored band). Peterson limits (NLNM in black dashed line and NHNM in black dotted line) and the self-noise of both sensors (black solid line) are also shown.}}
     \label{fig:seismicASDdisp}
 \end{figure}
 
Figure \ref{fig:seismicASDdisp} shows the comparison of the percentiles of the horizontal component for P2 (in blue) and TERZ (in red) with respect to the Peterson New Low and High Noise Model (NLNM in black dashed line and NHNM in black dotted line respectively) \cite{peterson1993observations}. Above 6 Hz, the P2 data lie on the self-noise of the sensor, making apparent the quietness of the site. We will use these data in Section \ref{sec:curves} as input parameters for the determination of the ET sensitivity curve.

\subsection{Astrophysical Sources}
\begin{figure}
     \centering
     \begin{subfigure}{0.7\columnwidth}
         \centering
\includegraphics[width=0.7\textwidth]{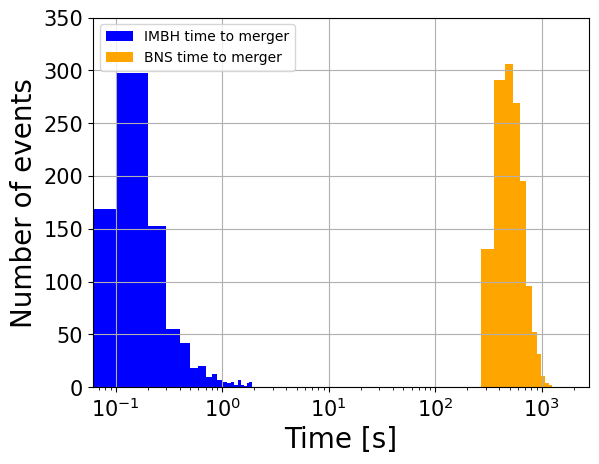}
         \caption{\textit{Distribution of the time to merger at $10\rm\, Hz$ for the events considered in this work.}}
         \label{fig:time_to_merge}
     \end{subfigure}
     \hfill
     \begin{subfigure}{0.7\columnwidth}
         \centering
         \includegraphics[width=0.7\textwidth]{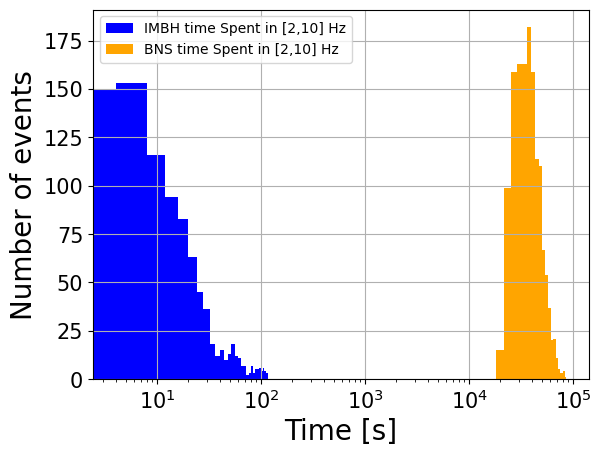}
         \caption{\textit{Distribution of the time spent in the $[2,10]\rm\,Hz$ range for the events considered in this work.}}
         \label{fig:time_spent}
     \end{subfigure}
        \caption{Times distributions for the events accounted for in this work.}
        \label{fig:bbh_times_in_range}
\end{figure}

The parameters for IMBH and BNS mergers are taken from the catalog \cite{cobacatalog} used to produce the results of \cite{coba}, which considers a set of sources spanning one year between the mock dates 01-01-2030 and 12-31-2030. For what concerns BBH, the catalog results from the mixture of BBH from isolated binary evolution, dynamical formation in young, globular and nuclear star clusters. Masses, spins, redshifts and luminosity distances have been obtained with the open-source code FASTCLUSTER \cite{fastcluster}. 
For this work, we select BBH with at least one source frame component mass above 100 $M_{\bigodot}$, for a total of $\simeq 10^3$ events. Figure \ref{fig:time_to_merge} (blue histogram) shows the distribution of the time to the merger, calculated numerically from the waveforms, at $10\rm\, Hz$ for these events. We note that once left the frequency band considered in this work, most events have less than $0.5\rm\, s$ before merging. On the other hand Figure \ref{fig:time_spent} (blue histogram) shows the time spent in the $[2,10]\rm\,Hz$ range and points out how most of the time IMBH lie in this bandwidth before the merger. 

For BNS mergers, the source frame masses of the two objects are sampled uniformly in the interval $[1.1; 2.5] \rm M_{\bigodot}$ so that $m_1>m_2$. Redshifts and luminosity distances are taken from \cite{santoliquido21}. Since \cite{cobacatalog} contains $\mathcal{O}( 7\times 10^5)$ BNS events, to limit the computational cost and time of the analysis we randomly sample, using a logarithmic distribution in redshift to privilege closer - therefore stronger - signals, $10^3$ events at $z < 1.5$. As mentioned, the BNS signal should reach $SNR \geq 12$ to issue an early warning between 20 hours and 1 hour before the merger. Figure \ref{fig:time_to_merge} (orange histogram) shows the distribution of the time to the merger at $10\rm\, Hz$ for the selected BNS events. We note that, once left the [2,10] Hz frequency band, most events have less than $20\rm\,{min}$ before merger. Figure \ref{fig:time_spent} (orange histogram) shows the time spent in the $[2,10]\rm\,Hz$ range. The relevance of this bandwidth is apparent, since several hours are spent here.

For both IMBH and BNS, the sky position and coalescence phase are sampled uniformly in the entire sky and between $[0, 2\pi]$, respectively. The inclination angle is sampled from a uniform distribution between $[-\pi,\pi]$ and the polarization angle is sampled uniformly in the interval $[0,\pi]$. For simplicity, we are also assuming the circular orbit case. Moreover, to take redshift into account, source frame masses are converted into the detector frame as $m_{det} = m_{sour}(1+z)$.
\section{Results}
\subsection{Modified ET curves}
\label{sec:curves}
\begin{figure}
     \centering
     \includegraphics[width=0.5\linewidth]{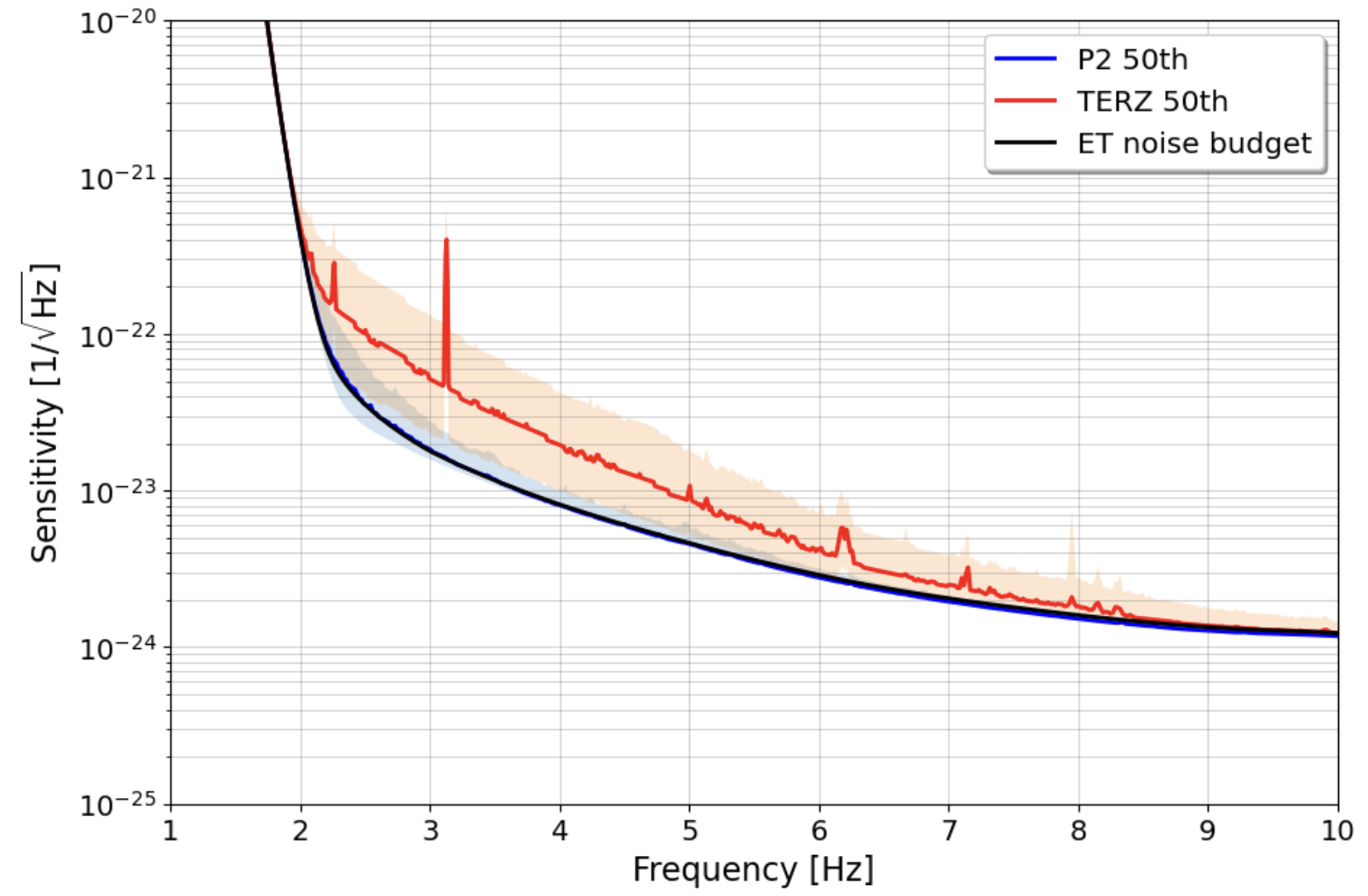}
     \caption{\textit{Sensitivity curves for the two candidate sites, reporting the percentile (50th in solid line while 10th and 90th the lower and upper limit of the colored band) for TERZ (in red) and P2 (in blue), compared with the ET design sensitivity (black).}}
     \label{fig:strainNN}
 \end{figure}

The ET sensitivity, modified according to what described in Section \ref{sec:Method}, takes into account the local seismic component of the NN of the two candidate sites. In equation \ref{eq:NN}, we assume $\rho_0=2.7\cdot10^3$ kg/m$^3$ and $L=10$ km. Figure \ref{fig:strainNN} reports the sensitivity curve using the NN from P2 and TERZ compared with the ET design noise budget as reported in \cite{coba} for the frequency range between 1 Hz and 100 Hz.

To quantify the modified ET sensitivity with respect to the designed one, we evaluated the NTR index defined in Sec. \ref{sec:Method}. While Fig. \ref{fig:strainNN} shows the level of the NN inside the ET sensitivity averaged over two years, the NTR, computed every $120\rm\,s$, gives information on how many times the modified sensitivity ($S_{n}(f)$) is above ($>1$) ore below ($<1$) the designed one ($S_{n,\rm{ET}}(f)$), i.e., information related to the glitchiness of the site. Focusing our attention on the frequency range between $f_1=2$ Hz and $f_2=10$ Hz, a value of NTR larger than one implies that, in the selected dataset, the contribution of the NN is limiting the ET sensitivity. The higher the NTR, the higher the probability of losing the GW signals. One has to remember that if the bandwidth of the GW signal is larger than $f_2$ it is possible to recover the signal since the NN is not significant at higher frequencies.

 \begin{figure}
     \centering
     \includegraphics[width=0.5\linewidth]{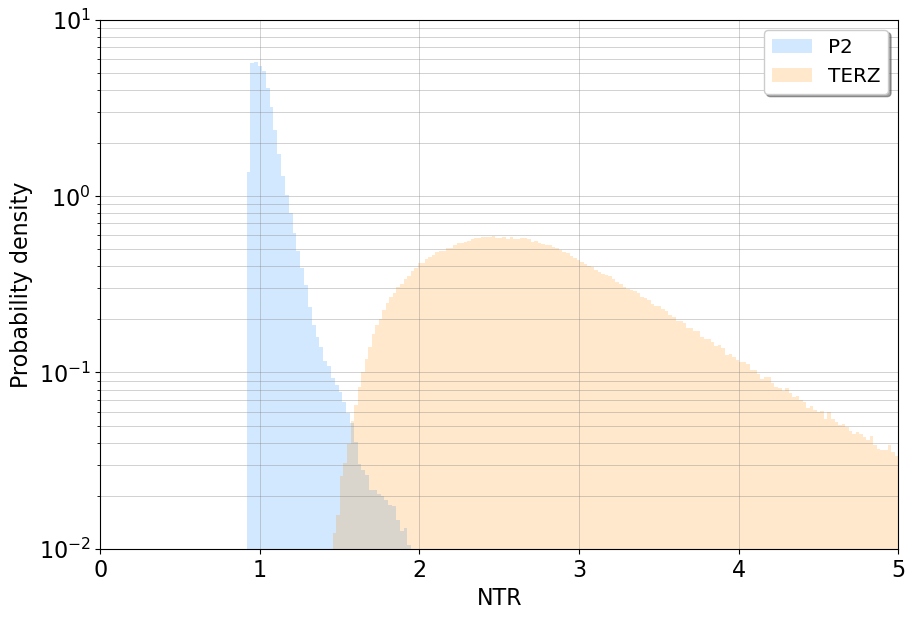}
     \caption{\textit{NTR normalised probability distribution for the two sites: P2 (in blue) and TERZ (in red).}}
     \label{fig:NTR}
 \end{figure}

\begin{figure}
     \centering
     \includegraphics[width=0.5\linewidth]{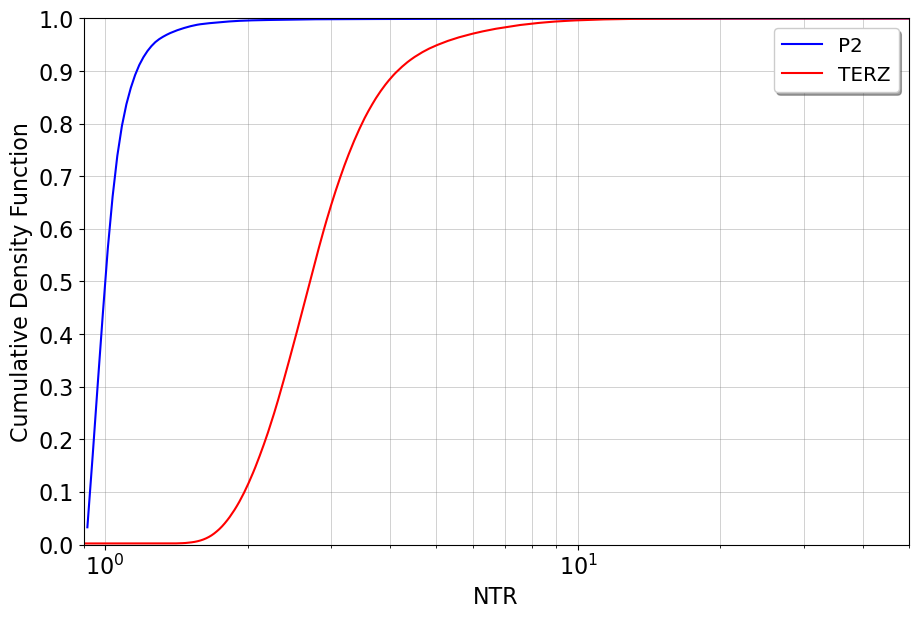}
     \caption{\textit{Normalised cumulative density function of the NTR normalised distribution, of Fig. \ref{fig:NTR}, for the two sites: P2 (in blue) and TERZ (in red).}}
     \label{fig:CDF}
 \end{figure}

Figure \ref{fig:NTR} reports the NTR normalised probability distribution for P2 and TERZ. It is possible to note that P2 is characterised by a lower noise and the majority of the data are distributed around $NTR=1$. To emphasize the difference, we evaluated the cumulative density function (CDF) of the two distributions. The x-axis of Fig. \ref{fig:CDF} is the NTR and the y-axis represents the CDF confined between 0 (no data are below the specific NTR) and 1 (all data are below that specific NTR). We reported in Table \ref{tab:NTR_CDF} the value of the NTR corresponding to the specific percentile for the two sites and in different time periods. The first set of data regards the whole period of two years, 90\% of the P2 data are below $NTR=1.31$, while the same amount of data for TERZ are below $NTR=4.69$. We performed the analysis separating working days and weekends, day and night hours. The noise is less on both weekends and nights giving information on the anthropogenic origin of this noise related to human activities. Moreover, this human noise is affecting more the TERZ site, while for P2 the variation is less significant.

\begin{table*}
\centering
\begin{tabular}{ |p{0.08\textwidth}||p{0.05\textwidth}|p{0.08\textwidth}|p{0.05\textwidth}|p{0.08\textwidth}|p{0.05\textwidth}|p{0.08\textwidth}|p{0.05\textwidth}|p{0.08\textwidth}|p{0.05\textwidth}|p{0.08\textwidth}|}
 \hline
    & \multicolumn{2}{|c|}{\textbf{Whole Period}} & \multicolumn{2}{|c|}{\textbf{Working days}} & \multicolumn{2}{|c|}{\textbf{Weekends}} & \multicolumn{2}{|c|}{\textbf{Days}} & \multicolumn{2}{|c|}{\textbf{Nights}} \\
 \hline
   \textbf{Perc.} & \textbf{P2} & \textbf{TERZ} & \textbf{P2} & \textbf{TERZ} & \textbf{P2} & \textbf{TERZ} & \textbf{P2} & \textbf{TERZ} & \textbf{P2} & \textbf{TERZ} \\
 \hline
 10th & $0.94$ & $1.59$ & $0.94$ & $1.66$ & $0.94$ & $1.49$ & $0.96$ & $1.84$ & $0.93$ & $1.44$ \\
 50th & $1.06$ & $2.49$ & $1.08$ & $2.68$ & $1.04$ & $2.11$ & $1.09$ & $2.84$ & $1.01$ & $1.86$ \\
 90th & $1.31$ & $4.69$ & $1.33$ & $4.96$ & $1.25$ & $3.81$ & $1.34$ & $4.94$ & $1.20$ & $3.68$ \\
 \hline
\end{tabular}
\caption{\textit{NTR values at the specified percentile (10th, 50th, and 90th) for both sites and in different conditions (whole period of two years, working day, weekends, days, nights).}}
\label{tab:NTR_CDF}
\end{table*}

\subsection{Impact on Signal to Noise Ratio}
\label{sec:snr}
\subsubsection{Intermediate Mass Black Holes}\label{sec:imbh}
\begin{figure}[p]
\centering
    \begin{subfigure}[]{0.7\columnwidth}
         \centering
         \includegraphics[width=0.7\textwidth]{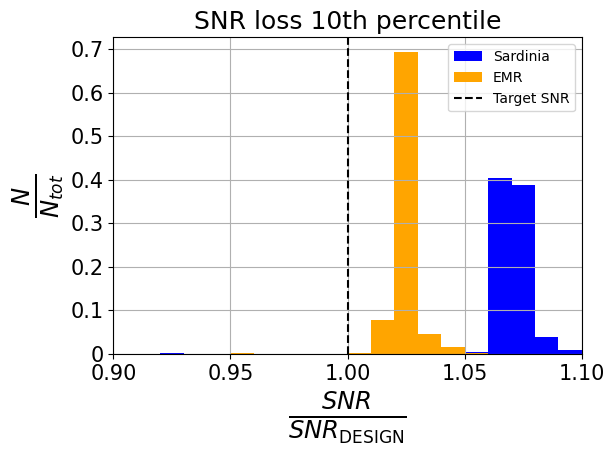}
         \caption{\textit{Sardinia SNR/SNR$_{\rm{design}}$ using the 10th percentile ET sensitivity.}}
         \label{fig:snrd1}
     \end{subfigure}
     \hfill
     \begin{subfigure}[]{0.7\columnwidth}
         \centering
         \includegraphics[width=0.7\textwidth]{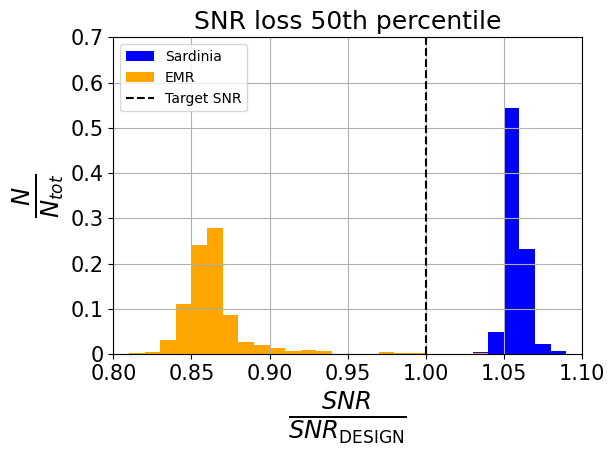}
         \caption{\textit{Sardinia SNR/SNR$_{\rm{design}}$ using the 50th percentile ET sensitivity.}}
         \label{fig:snre1}
     \end{subfigure}
     \hfill
     \begin{subfigure}[]{0.7\columnwidth}
         \centering
         \includegraphics[width=0.7\textwidth]{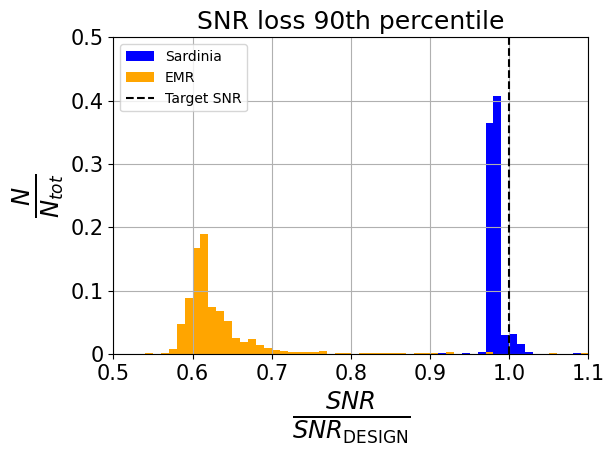}
         \caption{\textit{Sardinia SNR/SNR$_{\rm{design}}$ using the 90th percentile ET sensitivity.}}
         \label{fig:snrf1}
     \end{subfigure}
        \caption{\textit{IMBH SNR loss distributions for the Sardinia (blue) and EMR (orange) candidate sites for the 10th, 50th and 90th percentile of the seismic noise. The target SNR is shown as a vertical black dashed line.}}
        \label{fig:snr_dist_sos}
\end{figure}

Figure \ref{fig:snr_dist_sos} shows the distributions of the ratio between the measured SNR and the SNR calculated for the design sensitivity case ($\frac{SNR}{SNR_{\rm{DESIGN}}}$) for the IMBH case for the Sardinia and EMR candidate sites, respectively. These distributions show any loss/gain of SNR with respect to the design sensitivity case. The distributions of the SNR after which the $\frac{SNR}{SNR_{\rm{DESIGN}}}$ ratio is inferred are reported in the Appendix. The overall results, including all classes of astrophysical objects considered in this work, are also summarized in Table \ref{tab:imbh_summary}. For both Sardinia and EMR cases, we analyze the same 970 IMBH events and we exclude all those events which have $\rm SNR < 12$ in the best noise condition already. These events never reach the SNR = 12 threshold at 10 Hz also when the noise worsens and therefore are not relevant for the scope of this work. On the contrary, this cut guarantees that events eventually lost in the design case, but recovered in the best noise conditions, are retained. This can also highlight any possible improvement of the performances with respect to the expectations. 

In Sardinia, on average, the 10th and 50th percentile cases show comparable results with an SNR gain of $\simeq 7 \%$ and $\simeq 6 \%$, respectively (Figures \ref{fig:snrd1} and \ref{fig:snre1}, blue histogram).  On the other hand, in the 90th percentile case the performance is on par with the design case: there is only a marginal SNR loss ($\simeq 2.5 \%$ - Figure \ref{fig:snrf1}, blue histogram).

For the EMR site, the situation is different. The 10th percentile case is on par with the design performance, with the SNR showing a gain of $\simeq 2.5 \%$ (Figure \ref{fig:snrd1}, orange histogram). In the 50th percentile, the SNR distribution starts do deviate from the design case, with a SNR loss $\simeq 14 \%$ (Figure \ref{fig:snre1}, orange histogram). The 90th percentile shows, as expected, the worst performance with an average SNR loss $\simeq 37\%$ (Figure\ref{fig:snrf1}, orange histogram). A summary of the results is reported in Table \ref{tab:imbh_summary}.
\begin{table*}[t]
\centering
\begin{tabular}{ |p{0.08\textwidth}||p{0.1\textwidth}|p{0.1\textwidth}|p{0.15\textwidth}| p{0.15\textwidth}| p{0.23\textwidth}| }
 \hline
    & \multicolumn{2}{|c|}{\textbf{SNR/SNR$\rm\,_{DESIGN}$}} & \multicolumn{2}{|c|}{\textbf{EVENTS WITH SNR $< 12$}} & \textbf{SNR$_{\rm{EMR}}$/SNR$_{\rm{SAR}}$} \\
 \hline
   \textbf{Perc.} & \textbf{Sardinia} & \textbf{EMR} & \textbf{Sardinia} & \textbf{EMR} & \\
 \hline
 10th & $+7\%$ &$+2.5\%$&$-2\%$& $-1\%$&$0.96$\\
 50th & $+5\%$ &$-13\%$& $-1.8\%$ & $5\%$&$0.83$\\
 90th &  $-2.5\%$ &$-37\%$& $2.5\%$ & $24\%$&$0.69$\\
 \hline
\end{tabular}
\caption{\textit{Summary of the SNR performance for the Sardinia and EMR candidate sites. The figures reported in this table are the average of the total IMBH and BNS cases. The separate tables for the two classes of events are reported in the Appendix. The negative figures under the events with SNR $<12$ highlight an improvement in the fraction of recovered events with respect to the design case.}}
\label{tab:imbh_summary}
\end{table*}
\begin{figure}[p]
\centering
     \begin{subfigure}[]{0.5\columnwidth}
         \centering
         \includegraphics[width=\textwidth]{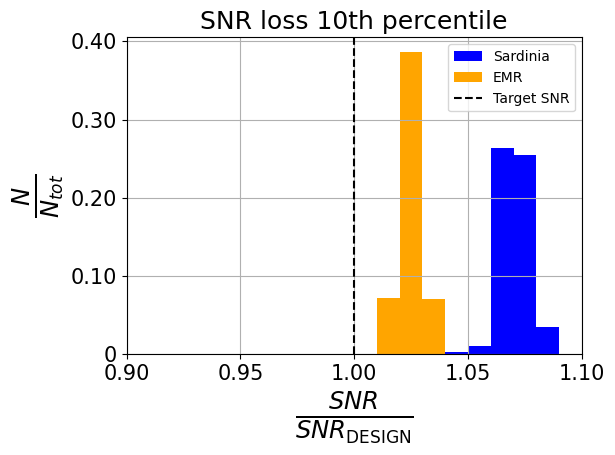}
         \caption{\textit{Sardinia SNR/SNR$_{\rm{design}}$ using the 10th percentile ET sensitivity.}}
         \label{fig:BNSsnrd1}
     \end{subfigure}
     \hfill
     \begin{subfigure}[]{0.5\columnwidth}
         \centering
         \includegraphics[width=\textwidth]{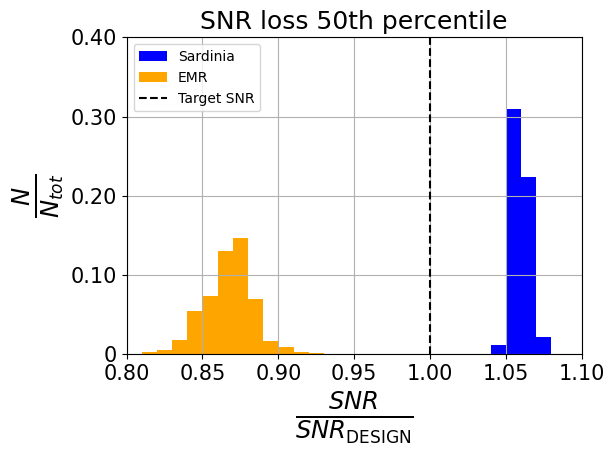}
         \caption{\textit{Sardinia SNR/SNR$_{\rm{design}}$ using the 50th percentile ET sensitivity.}}
         \label{fig:BNSsnre1}
     \end{subfigure}
     \hfill
     \begin{subfigure}[]{0.5\columnwidth}
         \centering
         \includegraphics[width=\textwidth]{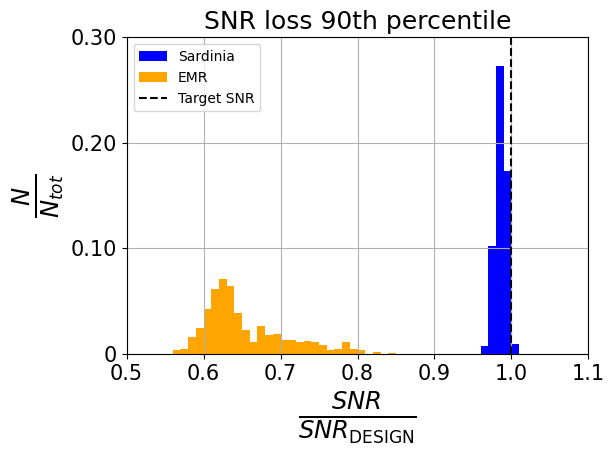}
         \caption{\textit{Sardinia SNR/SNR$_{\rm{design}}$ using the 90th percentile ET sensitivity.}}
         \label{fig:BNSsnrf1}
     \end{subfigure}
        \caption\textit{{BNS SNR loss distributions for the Sardinia (blue) and EMR (orange) candidate sites for the 10th, 50th and 90th percentile of the seismic noise. The target SNR is shown as a vertical black dashed line.}}
        \label{fig:BNSsnr_dist_sos}
\end{figure}

\begin{figure}[p]
     \centering
     \begin{subfigure}[]{0.5\columnwidth}
         \centering
         \includegraphics[width=\textwidth]{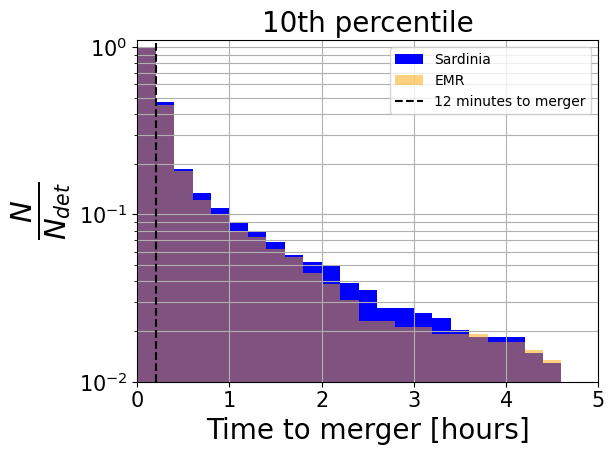}
         \caption{}
         \label{fig:BNStimea5}
     \end{subfigure}
     \hfill
     \begin{subfigure}[]{0.5\columnwidth}
         \centering
         \includegraphics[width=\textwidth]{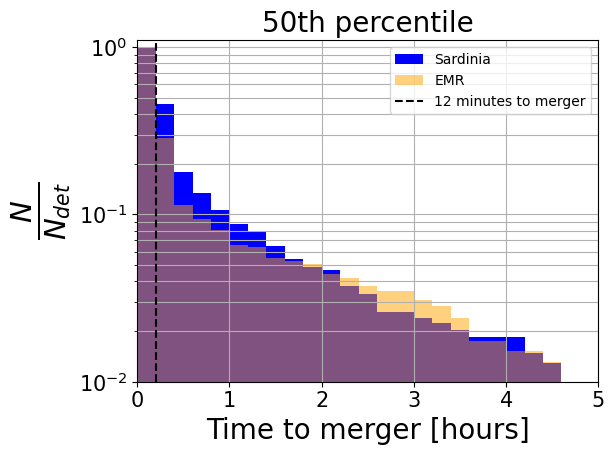}
         \caption{}
         \label{fig:BNStimeb5}
     \end{subfigure}
    \hfill
     \begin{subfigure}[]{0.5\columnwidth}
         \centering
         \includegraphics[width=\textwidth]{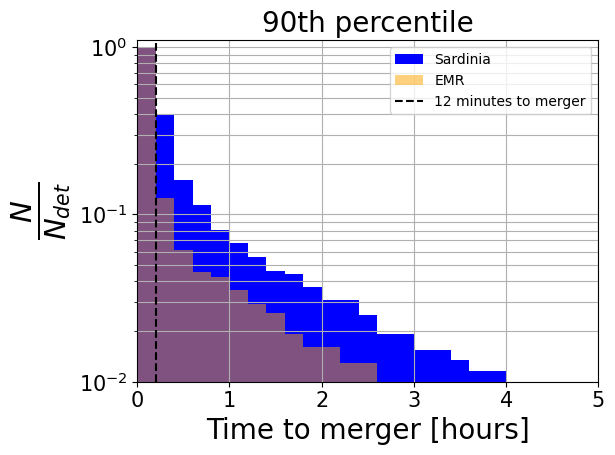}
         \caption{}
         \label{fig:BNStimec5}
     \end{subfigure}
        \caption{Cumulative distributions of the time to merger at detection for the Sardinia (blue) and EMR (orange) sites.}
        \label{fig:BNStime_to_dist_sos}
\end{figure}
\subsubsection{Binary Neutron Stars}

Figure \ref{fig:BNSsnr_dist_sos} shows the results obtained in the case of BNS signals. As for the previous results, we exclude events with SNR $<12$ in the best noise case. 
For Sardinia, the results are consistent with the IMBH analysis and, in all cases, the ratios between the reconstructed SNR and the design SNR show only marginal differences with respect to what discussed in the previous section (Figure \ref{fig:BNSsnr_dist_sos}, blue histograms).

The EMR site shows again the less promising performances. Again (Figure \ref{fig:snr_dist_sos}, orange histograms) the BNS case show results which are consistent with the IMBH case, with the 10th percentile case equaling the figure of the benchmark case, with a SNR gain $\simeq 2\%$, and the 50th and 90th percentile showing the worst performances loosing $\simeq 14\%$ and $\simeq 37\%$ of the SNR, respectively.

Since the time to merger at detection is a relevant information for multimessenger studies, we also provide the cumulative distributions of the fraction of detected events vs. the time to merger at the moment in which the threshold SNR = 12 is reached (Figure \ref{fig:BNStime_to_dist_sos}). The estimation of the time to merger is done numerically from the simulated waveforms, i.e., we mark the time at which each event reaches the SNR threshold and infer from the waveform how far it is from the merger time.

At the EMR site, we notice that in the worst possible scenario only $11\%$ of events are detectable within 12 minutes from merger, whereas, in the best noise conditions, this fraction improves to $45\%$. For the 50th percentile case, on the other hand, the fraction is $30\%$. In Sardinia the difference between the three cases is less apparent, since the fraction of events detectable within 12 minutes from merger oscillates between $40\%$ and $47\%$.

\section{Discussion}
\label{sec:disc}
We investigated the impact of site dependent ambient noise on the low-frequency sensitivity of ET at the two sites candidate to host ET: the Sos Enattos site in Sardinia, Italy, and the Terziet side in the EMR region at the border between The Netherlands and Belgium. According to the currently available site characterization data, we focused on the impact of the seismic noise contribution to NN which is more prominent between $2 \rm\, Hz$ and $10 \rm\, Hz$. Using the 10th, 50th and 90th percentile of the seismic spectra distributions covering two years at the two candidate sites, we inferred the impact on the ET sensitivity curve. We found that the Sardinia candidate site has the lowest impact on the ET sensitivity if compared to the effect of local ambient noise recorded at EMR, which, on the other hand, appears to have a larger impact. These results were obtained without considering any NN suppression factor in Equation \ref{eq:NN}. As a consequence, we expect that these results may provide some relevant information about the design of NN suppression systems for ET that will be different for each candidate site. In fact, Sardinia appears to be compliant with the requirements of ET already without considering any NN suppression factor, whereas the larger impact on the design sensitivity of the EMR site will require dedicated seismometer arrays for noise subtraction \cite{vanBeveren_2023} to reduce its impact on the design sensitivity. These results suggests that Sardinia is also an ideal location for research and development activities in low noise environments. For what concerns achievable NN suppression factors, previous works \cite{badaracco2019, Badaracco2020, Koley2024} showed that by optimizing NN cancellation for a specific frequency may lead up to a factor 10 suppression. Nevertheless, at other frequencies the factor would be $\simeq 2$ \cite{Badaracco2020}. Broadband NN cancellation, on the other hand, would not achieved a suppression factor better than 3 \cite{Badaracco2020}. It should also be pointed out that the signals considered in this work span a wide range of frequencies, therefore we should focus on broadband NN cancellation. Finally, NN cancellation comes at the cost of a huge computational and financial effort. For example,  considering the achievement of a factor 3 NN suppression factor as the most reasonable assumption for ET, it would require a few tens of seismometers in boreholes per test mass \cite{Badaracco2020}. Since the ET triangle will have 12 test masses, we should expect $\mathcal{O}(100)$ sensors placed in boreholes around the test masses. The main challenges would be determining where to drill the boreholes to achieve effective NN reduction and the processing of 3D data for the NN cancellation algorithms (being on surface, current detectors only require 2D data).

The different impact of the Sardinia and EMR sites on the ET sensitivity translates into an influence on the signal SNR at low frequency for those astrophysical sources for which low-frequency sensitivity plays an important role, i.e. IMBH and BNS mergers. Using the source parameters present in astrophysical catalogs obtained from population studies, we generated a set of waveforms and injected them into ET noise simulated according to the sensitivity curves derived from local ambient noise. SNR has then been calculated using the matched filter SNR. Comparing the resulting SNR distributions against the benchmark case obtained from the ET design sensitivity curve, we find that local ambient noise has a lower impact on the signal SNR in Sardinia, which is very close to design. The EMR site shows a higher SNR degradation, with respect to design (Figures \ref{fig:snr_dist_sos} and \ref{fig:BNSsnr_dist_sos}).
\section{Conclusions}
\label{sec:conc}
 Given the ET early warning requirements for BNS and the time to merger at $10 \rm\, Hz$ of these signals, a significant degradation of the signal SNR may affect the ET early warning capabilities and hinder the possibility of extensive multimessenger observation campaigns for a number of events larger than expected. In this sense, we have also inferred the cumulative distributions of the time to merger at detection and found that this is most likely to happen at EMR than Sardinia (Figure \ref{fig:BNStime_to_dist_sos}). These issues could be solved only with the use of appropriate NN suppression that should be more prominent at the EMR site.

In future, we will follow up this study to consider the ET 2L configuration and assess the impact of site dependent noise on a detector network composed of two widely separated detectors that are not affected by the same noise sources at the same times. Perhaps, in this case, the
presence of the most sensitive interferometer in Sardinia could make the impact of the worse
EMR sensitivity on the detection rates less dramatic. We also plan to quantify the effect of local ambient noise on the source sky localization areas issued by early warnings using currently available detection pipelines. This will further expand the overview of the impact of ambient noise on the detectability of GW sources.

\ack

The authors gratefully acknowledge dr. M. Mancarella, prof. T. Bulik, dr. S. Koley and dr. J. van Heijningen for their helpful comments and advices that significantly improved the presentation of this work. Dr. M. Di Giovanni would also like to acknowledge dr. E. Codazzo for providing the necessary information to correctly download and interpret the information contained in the astrophysical catalogs. This work is also partially funded by the PRIN 2020 PE9 project 2020BRP57Z of the Ministero dell'Università e della Ricerca.

\providecommand{\newblock}{}


\newcommand{\hbAppendixPrefix}{A}
\renewcommand{\thefigure}{\hbAppendixPrefix\arabic{figure}}
\setcounter{figure}{0}

\renewcommand{\thetable}{A\arabic{table}}
\setcounter{table}{0}

\section*{Appendix}
For completeness, in the Appendix we report the SNR distributions for the Sardinia and EMR sites for both the IMBH and BNS cases (Figures A1-A4). These distributions are those used to calculate the $\frac{SNR}{SNR_{\rm DESIGN}}$ ratio of Figures \ref{fig:snr_dist_sos} and \ref{fig:BNSsnr_dist_sos}. In the various figures, the reference distribution of the design sensitivity case is shown in green. As mentioned in the paper, we exclude from the analysis all the events that never reach SNR = 12 below 10 Hz in the 10th percentile case, since they will be never detected in any other case as well. This  also implies that some events recovered in the 10th percentile case (best possible noise conditions) may be missed in the benchmark case, since the 10th percentile sensitivity curve may be, at some frequencies, better than the design one. This is indicated by the green histogram bins appearing below the SNR = 12 threshold.

We also include the tables (Tables \ref{tab:appimbh_summary} and \ref{tab:appBNS_summary}) which contain the results for the IMBH and BNS cases separately that are summarized in Table \ref{tab:imbh_summary}.

\begin{figure*}[h]
     \centering
     \begin{subfigure}[b]{0.3\columnwidth}
         \centering
         \includegraphics[width=\textwidth]{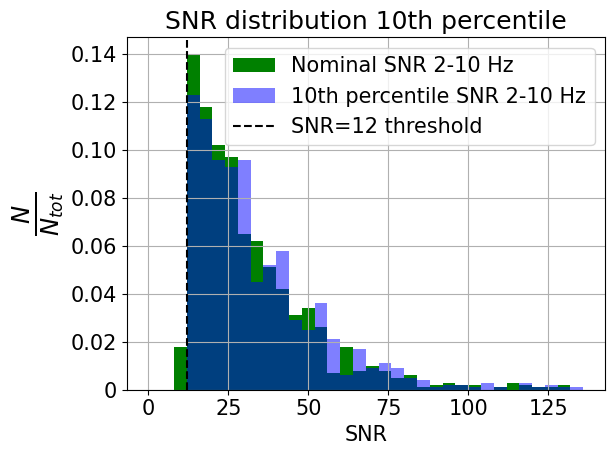}
         \caption{\textit{Sardinia Nominal SNR distribution (green) and SNR distribution using the 10th percentile ET sensitivity (blue).}}
         \label{fig:snra1}
     \end{subfigure}
     \hfill
     \begin{subfigure}[b]{0.3\columnwidth}
         \centering
         \includegraphics[width=\textwidth]{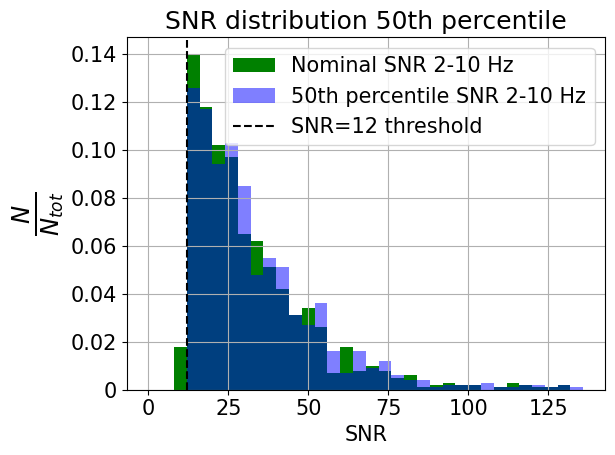}
         \caption{\textit{Sardinia nominal SNR distribution (green) and SNR distribution using the 50th percentile ET sensitivity (blue).}}
         \label{fig:snrb1}
     \end{subfigure}
    \hfill
     \begin{subfigure}[b]{0.3\columnwidth}
         \centering
         \includegraphics[width=\textwidth]{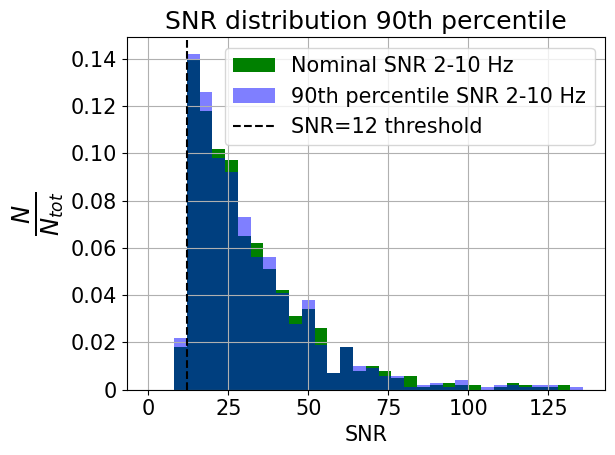}
         \caption{\textit{Sardinia nominal SNR distribution (green) and SNR distribution using the 90th percentile ET sensitivity (blue).}}
         \label{fig:snrc1}
     \end{subfigure}
        \caption{\textit{IMBH SNR distributions for the Sardinia candidate site. In the top row, the SNR = 12 threshold is marked with a black dashed line. In the bottom row, the black dashed line represents the equality between the SNR measured in the design case with the cases of the modified ET curves. The non overlapped green SNR bin under the SNR threshold highlights the fact that there are some events lost in the design case but recovered in the best possible recorded noise conditions.}}
        \label{fig:snr_dist_sos2}
\end{figure*}
\begin{figure*}[h]
     \centering
     \begin{subfigure}[b]{0.3\columnwidth}
         \centering
         \includegraphics[width=\textwidth]{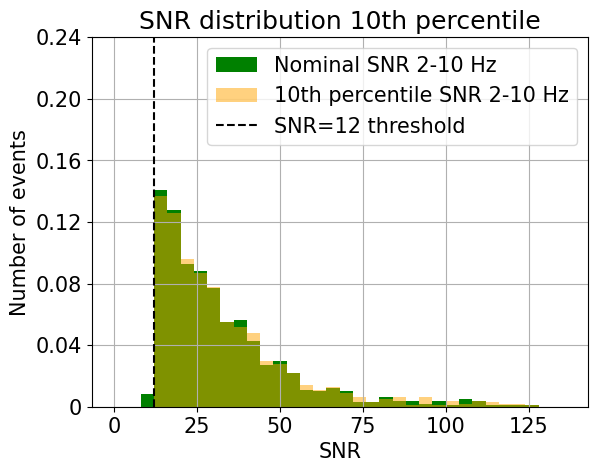}
         \caption{\textit{EMR Nominal SNR distribution (green) and SNR distribution using the 10th percentile ET sensitivity (blue).}}
         \label{fig:snra2}
     \end{subfigure}
     \hfill
     \begin{subfigure}[b]{0.3\columnwidth}
         \centering
         \includegraphics[width=\textwidth]{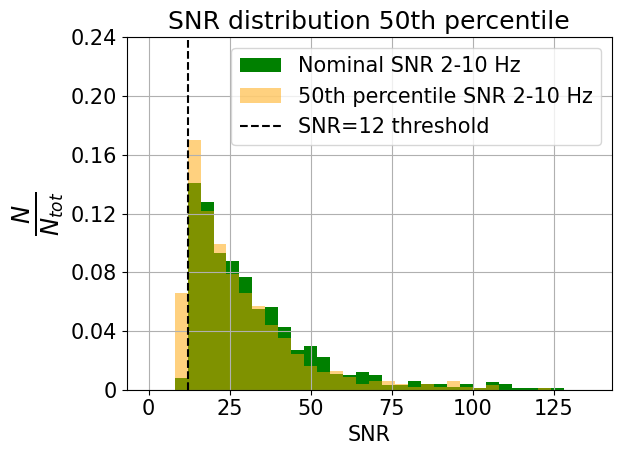}
         \caption{\textit{EMR Nominal SNR distribution (green) and SNR distribution using the 50th percentile ET sensitivity (blue).}}
         \label{fig:snrb2}
     \end{subfigure}
    \hfill
     \begin{subfigure}[b]{0.3\columnwidth}
         \centering
         \includegraphics[width=\textwidth]{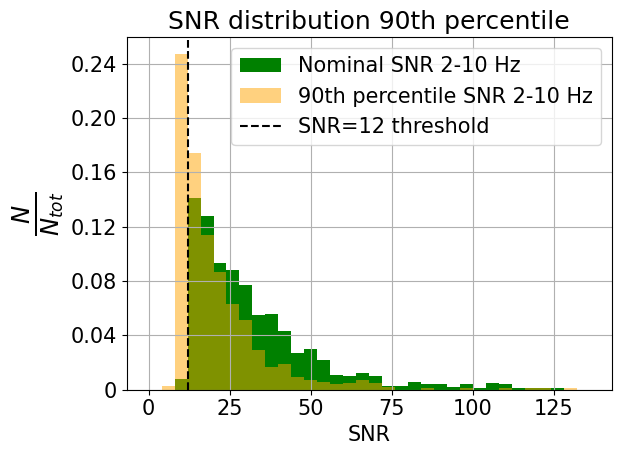}
         \caption{\textit{EMR Nominal SNR distribution (green) and SNR distribution using the 90th percentile ET sensitivity (blue).}}
         \label{fig:snrc2}
     \end{subfigure}
        \caption{\textit{IMBH SNR distributions for the EMR candidate site. In the top row, the SNR = 12 threshold is marked with a black dashed line. In the bottom row, the black dashed line represents the equality between the SNR measured in the design case with the cases of the modified ET curves. The non overlapped green SNR bin under the SNR threshold highlights the fact that there are some events lost in the design case but recovered in the best possible recorded noise conditions.}}
        \label{fig:snr_dist_terz2}
\end{figure*}
\begin{figure*}[h]
     \centering
     \begin{subfigure}[b]{0.3\columnwidth}
         \centering
         \includegraphics[width=\textwidth]{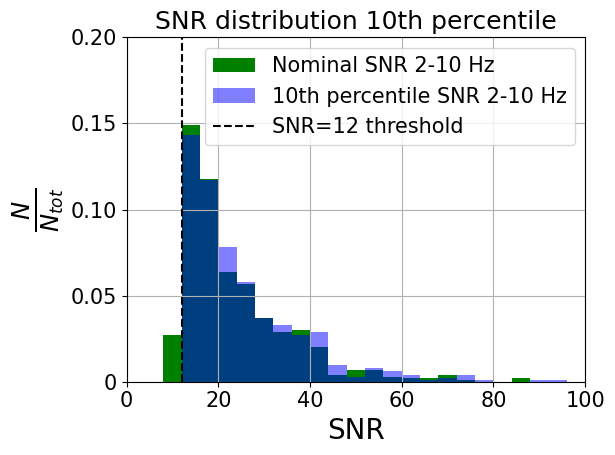}
         \caption{\textit{Sardinia Nominal SNR distribution (green) and SNR distribution using the 10th percentile ET sensitivity (blue).}}
         \label{fig:BNSsnra1}
     \end{subfigure}
     \hfill
     \begin{subfigure}[b]{0.3\columnwidth}
         \centering
         \includegraphics[width=\textwidth]{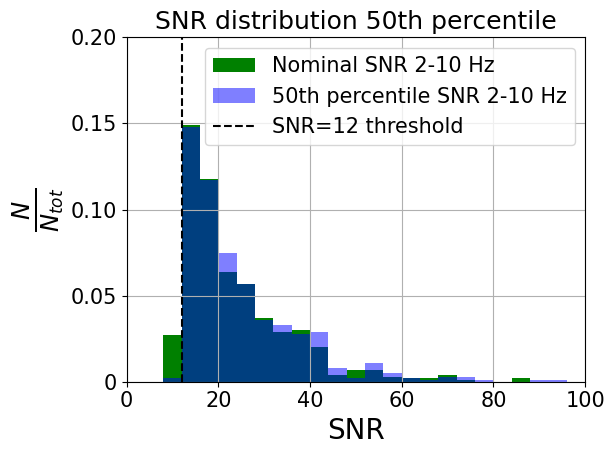}
         \caption{\textit{Sardinia nominal SNR distribution (green) and SNR distribution using the 50th percentile ET sensitivity (blue).}}
         \label{fig:BNSsnrb1}
     \end{subfigure}
    \hfill
     \begin{subfigure}[b]{0.3\columnwidth}
         \centering
         \includegraphics[width=\textwidth]{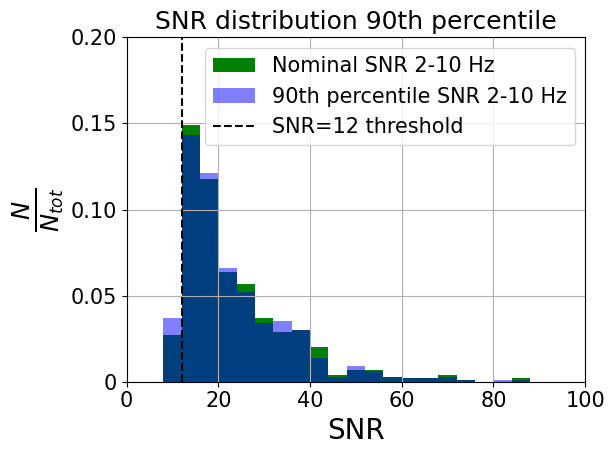}
         \caption{\textit{Sardinia nominal SNR distribution (green) and SNR distribution using the 90th percentile ET sensitivity (blue).}}
         \label{fig:BNSsnrc1}
     \end{subfigure}
        \caption\textit{{BNS SNR distributions for the Sardinia candidate site. In the top row, the SNR = 12 threshold is marked with a black dashed line. In the bottom row, the black dashed line represents the equality between the SNR measured in the design case with the cases of the modified ET curves. The non overlapped green SNR bin under the SNR threshold highlights the fact that there are some events lost in the design case but recovered in the best possible recorded noise conditions.}}
        \label{fig:BNSsnr_dist_sos2}
\end{figure*}
\begin{figure*}[h]
     \centering
     \begin{subfigure}[b]{0.3\columnwidth}
         \centering
         \includegraphics[width=\textwidth]{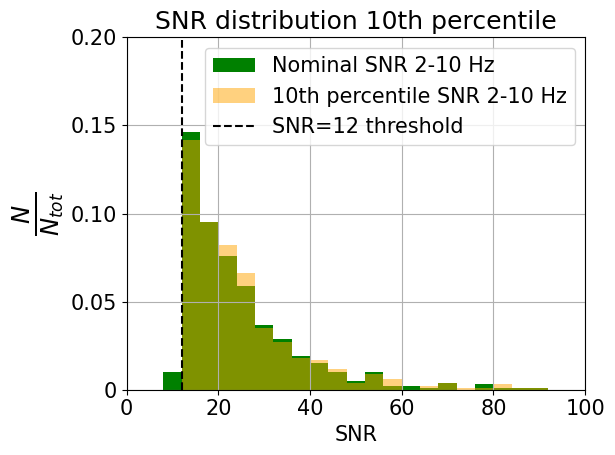}
         \caption{\textit{EMR Nominal SNR distribution (green) and SNR distribution using the 10th percentile ET sensitivity (blue).}}
         \label{fig:BNSsnra4}
     \end{subfigure}
     \hfill
     \begin{subfigure}[b]{0.3\columnwidth}
         \centering
         \includegraphics[width=\textwidth]{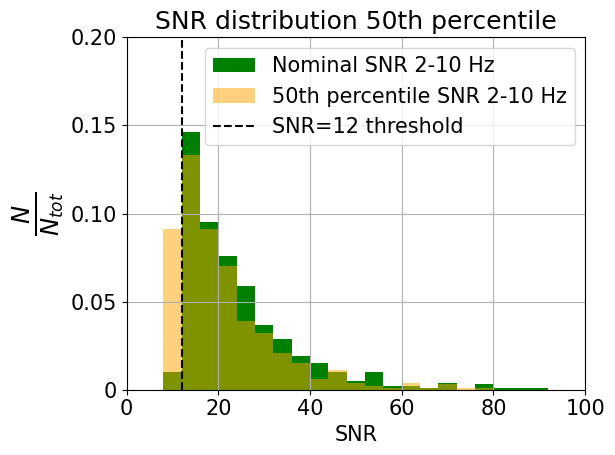}
         \caption{\textit{EMR Nominal SNR distribution (green) and SNR distribution using the 50th percentile ET sensitivity (blue).}}
         \label{fig:BNSsnrb4}
     \end{subfigure}
    \hfill
     \begin{subfigure}[b]{0.3\columnwidth}
         \centering
         \includegraphics[width=\textwidth]{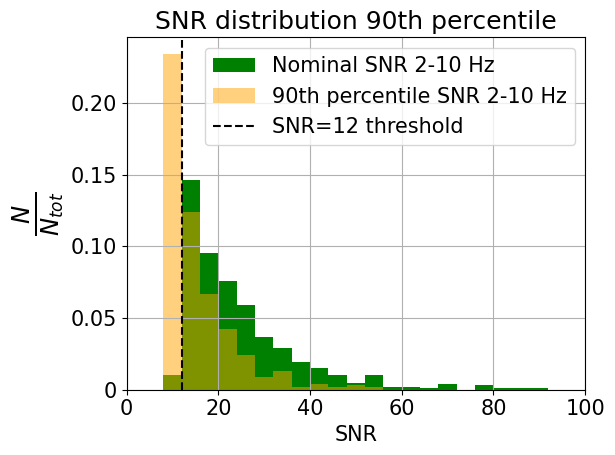}
         \caption{\textit{EMR Nominal SNR distribution (green) and SNR distribution using the 90th percentile ET sensitivity (blue).}}
         \label{fig:BNSsnrc4}
     \end{subfigure}
        \caption{\textit{BNS SNR distributions for the EMR candidate site. In the top row, the SNR = 12 threshold is marked with a black dashed line. In the bottom row, the black dashed line represents the equality between the SNR measured in the design case with the cases of the modified ET curves. The non overlapped green SNR bin under the SNR threshold highlights the fact that there are some events lost in the design case but recovered in the best possible recorded noise conditions.}}
        \label{fig:BNSsnr_dist_terz2}
\end{figure*}

\begin{table*}[h]
\centering
\begin{tabular}{ |p{0.08\textwidth}||p{0.15\textwidth}|p{0.15\textwidth}|p{0.15\textwidth}| p{0.15\textwidth}| p{0.25\textwidth}| }
 \hline
    & \multicolumn{2}{|c|}{\textbf{SNR/SNR$\rm\,_{DESIGN}$}} & \multicolumn{2}{|c|}{\textbf{EVENTS WITH SNR $< 12$}} & \textbf{SNR$_{EMR}$/SNR$_{SAR}$} \\
 \hline
   \textbf{Perc.} & \textbf{Sardinia} & \textbf{EMR} & \textbf{Sardinia} & \textbf{EMR} & \\
 \hline
 10th & $+7\%$ &$+2.5\%$&$-2\%$& $-1\%$&$0.96$\\
 50th   & $+6\%$ &$-14\%$& $-2\%$ & $6\%$&$0.83$\\
 90th &  $-2.5\%$ &$-37\%$& $2\%$ & $24\%$&$0.69$\\
 \hline
\end{tabular}
\caption{\textit{Summary of the SNR performance for the Sardinia and EMR candidate sites in the case of IMBH signals. The negative figures under the events with SNR $<12$ highlight an improvement in the fraction of recovered events with respect to the design case.}}
\label{tab:appimbh_summary}
\end{table*}

\begin{table*}[h]
\centering
\begin{tabular}{ |p{0.08\textwidth}||p{0.15\textwidth}|p{0.15\textwidth}|p{0.15\textwidth}| p{0.15\textwidth}| p{0.25\textwidth}| }
 \hline
    & \multicolumn{2}{|c|}{\textbf{SNR/SNR$\rm\,_{DESIGN}$}} & \multicolumn{2}{|c|}{\textbf{EVENTS WITH SNR $< 12$}} & \textbf{SNR$_{EMR}$/SNR$_{SAR}$} \\
 \hline
   \textbf{Perc.} & \textbf{Sardinia} & \textbf{EMR} & \textbf{Sardinia} & \textbf{EMR} & \\
 \hline
 10th & $+7\%$ &$+2.5\%$& $-2\%$ &$-1\%$&0.96\\
 50th   & $+5\%$ &$-12\%$& $-1.5\%$&$9\%$&0.83\\
 90th &  $-2.5\%$ &$-37\%$& $3\%$&$24\%$&0.69\\
 \hline
\end{tabular}
\caption{\textit{Summary of the SNR performance for the Sardinia and EMR candidate sites in the case of BNS mergers. The negative figures under the events with SNR $<12$ highlight an improvement in the fraction of recovered events with respect to the design case.}}
\label{tab:appBNS_summary}
\end{table*}

\end{document}